\documentclass[12pt]{article}

\usepackage{graphicx}
\usepackage{a4}

\usepackage{fancybox}
\usepackage{epsfig}
\usepackage{color}

\usepackage{amsfonts}
\usepackage{mathrsfs}
\usepackage{amssymb}
\usepackage{amsmath}

\usepackage{dcolumn}
\usepackage{bm}

\def\pa{\partial}

\def\th{\theta}

\def\ka{\kappa}

\def\om{\omega}

\def\Ga{\Gamma}

\def\Om{\Omega}

\newcommand{\ben}{\begin{equation}}
\newcommand{\een}{\end{equation}}
\newcommand{\bea}{\begin{eqnarray}}
\newcommand{\eea}{\end{eqnarray}}
\newcommand{\ba}{\begin{array}}
\newcommand{\ea}{\end{array}}
\newcommand{\bit}{\begin{itemize}}
\newcommand{\eit}{\end{itemize}}

\newcommand{\vs}[1]{\vspace{#1 mm}}

\newcommand{\dsl}{\pa \kern-0.5em /}

\begin{document}

\topmargin 0pt \oddsidemargin 0mm

\vspace{2mm}

\begin{center}

{\Large Hawking-like radiation of charged particles via tunneling
across the lightcylinder of a rotating magnetosphere}

\vs{10}

 {\large Huiquan Li \footnote{E-mail: lhq@ynao.ac.cn}}

\vspace{6mm}

{\em

Yunnan Observatories, Chinese Academy of Sciences, \\
650216 Kunming, China

Key Laboratory for the Structure and Evolution of Celestial Objects,
\\ Chinese Academy of Sciences, 650216 Kunming, China

Center for Astronomical Mega-Science, Chinese Academy of Sciences,
\\ 100012 Beijing, China}

\end{center}

\vs{9}

\begin{abstract}

In rotating magnetospheres planted on compact objects, there usually
exist lightcylinders (LC), beyond which the rotation speed of the
magnetic field lines exceeds the speed of light. The LC is a close
analog to the horizon in gravity, and is a casual boundary for
charged particles that are restricted to move along the magnetic
field lines. In this work, it is proposed that there should be
Hawking-like radiation of charged particles from the LC of a
rotating magnetosphere from the point of view of tunneling by using
the field sheet metric.

\end{abstract}



\section{Introduction}
\label{sec:introduction}

Particles can be created and radiated from the horizons of black
holes and the universe \cite{Hawking:1974sw,Gibbons:1977mu}. Outside
the horizon of a black hole, virtual particle pairs are excited from
the vacuum. The positive energy particle may escape from the near
horizon region, forming Hawking radiation, while the negative energy
particle falls into the black hole. The Hawking radiation from a
black hole can also be interpreted as particles tunneling across the
horizon \cite{Parikh:1999mf}. When particle pairs are created inside
the horizon, the positive energy particles can tunnel through the
horizon and come out, which is forbidden by the classical theory.
When pairs are created outside the horizon, the negative energy
particles tunnel from outside to the interior.

Analogous horizon also exists in electrodynamics. In constructing
magnetospheric theory on rotating and compact objects, we usually
encounters a critical surface, called the lightcylinder (LC). Beyond
the LC, the rotation speed of the magnetic field lines exceeds the
speed of light. In pulsar magentospheres, the LC is a turning
surface of geometry and topology of the field lines. It is a horizon
for charged particles and Alfv\'{e}n waves propagating along the
magnetic field lines. Once a charged particle moves out of the LC,
it will never cross the LC and get inside again. This is quite
similar to the case of horizons in gravity.

When a pair of charged particles are excited just inside the LC, one
particle may cross the LC and never comes back, left the other
particle inside the LC. Similarly, a charged particle outside the LC
may also tunnel across the LC and gets inside, left the other
particle remain outside the LC. Hence, there should exist radiation
of particles from the LC towards both sides. In this work, we shall
consider this by using the field sheet metric obtained in
\cite{2014MNRAS.445.2500G}. The field sheet is a two dimensional
``spacetime" that governs the motion of the charged particles that
move along the magnetic field lines. So here only charged particles
are relevant. This is different from the case of a horizon in
gravity, for which any particles that exist in nature can be
radiated.

\section{The field sheet of a rotating magnetosphere}
\label{sec:fieldsheet}

The magnetospheres on compact objects are usually filled with large
enough amount of charged particles so that the electric field
parallel to the magnetic field line is screened, i.e., the
electromagnetic fields satisfy the degenerate condition. Moreover,
the electromagnetic fields are very strong on these objects and the
inertia of charges can be neglected. So the force-free condition is
usually assumed.

As shown in
\cite{Carter:1979we,1997PhRvE..56.2181U,1997PhRvE..56.2198U,
2014MNRAS.445.2500G}, the degenerate electromagnetic field
satisfying the condition $F ^\star F=0$ can be expressed as
\begin{equation}
 F=d\phi_1\wedge d\phi_2,
\end{equation}
in terms of the two Euler scalars. In components, this means that
the electric field and magnetic field lines are perpendicular:
$\vec{E}\cdot\vec{B}=0$, which can be derived from the force-free
condition: $F_{\mu\nu}J^\nu=0$. The gauge field potential is given
by:
\begin{equation}
 A_\mu=\frac{1}{2}(\phi_1\pa_\mu\phi_2-\phi_2\pa_\mu\phi_1).
\end{equation}

For stationary and axisymmetric magnetospheres, the Euler potentials
take the following forms
\begin{equation}
 \phi_1=\psi(r,\th), \textrm{ }\textrm{ }\textrm{ }
\phi_2=\varphi-\Om t+f(r,\th).
\end{equation}
The functions $\psi$ and $f$ determines the poloidal current $I$
that flows along the magnetic field lines:
\begin{equation}
 F_{r\th}=\pa_r\psi\pa_\th f-\pa_rf\pa_\th\psi=\frac{I}{\sin\th}.
\end{equation}
In the force-free approximation, the angular velocity of the
magnetic field lines $\Om$ and the poloidal current $I$ are both
functions of $\psi$ generally. They satisfy the stream equation in
Minkowski spacetime for constant $\Om$:
\begin{eqnarray}\label{e:streameq}
 (1-r^2\sin^2\th\Om^2)\left(\pa_r^2\psi
+\frac{1}{r^2}\pa_\th^2\psi\right)
-2r\sin^2\th\Om^2\pa_r\psi-\frac{1}{r^2}(1+r^2\sin^2\th\Om^2)
\cot\th \pa_\th\psi=-II'.
\end{eqnarray}

The field sheet is defined as the intersection of the constant Euler
potentials. It is like the spacetime that governs the propagation of
the charged particles and Alfv\'{e}n waves on the field lines.
Substituting the conditions $d\phi_1=d\phi_2=0$ into the Minkowski
spacetime in spherical coordinates, we get the two-dimensional field
sheet metric:
\begin{equation}
 ds^2=-(1-r^2\sin^2\th\Om^2)dt^2+\frac{2r^2\sin\th\Om I}{\pa_\th\psi}dtdr
+\left[1+\frac{r^2((\pa_r\psi)^2+I^2)}{(\pa_\th\psi)^2}\right]dr^2.
\end{equation}
For vacuum solutions with $I=0$, the metric is static.

This two-dimensional metric generally has vanishing Einstein tensor:
$G_{ij}=T_{ij}^{\textrm{matter}}=0$, where $i,j$ denote the
coordinates $(t,r)$. This is analogous to the vacuum spacetime
around a black hole. It is consistent with the force-free
approximation, for which the inertial of the charged particles is
neglected.

\section{Tunneling across the LC}
\label{sec:tunneling}

An exact solution to the stream equation (\ref{e:streameq}) is the
monopole solution found in \cite{1973ApJ...180L.133M}:
$\psi=-q\cos\th$ and $I(\psi)=(\psi^2-q^2)/q$. The monopole profile
is thought to be the geometry of pulsar magnetospheres outside the
LC. For this solution, the field sheet metric is
\cite{2014MNRAS.445.2500G}
\begin{equation}
 ds^2=-(1-r^2\sin^2\th\Om^2)dt^2-2r^2\sin^2\th\Om^2dtdr
+(1+r^2\sin^2\th\Om^2)dr^2.
\end{equation}
This metric in the Eddington-Finkelstein coordinates looks like the
de Sitter metric, but is not exactly the same one. The horizon is
located at $r=1/(\Om \sin\th)$ with constant $\th$. The surface
gravity of the horizon is $\ka=\Om\sin\th$.

Following the standard procedure \cite{Parikh:1999mf}, we can
calculate the emission rate of particles due to the tunneling across
the LC by using the above metric. Here the tunneling calculation is
similar to the case for de Sitter space
\cite{Parikh:2002qh,Medved:2002zj}. The transmission coefficient for
crossing the classically forbidden region is given by
\begin{equation}
 \Ga=e^{-\frac{2}{\hbar} \textrm{ Im } S}.
\end{equation}

The equation of motion along the null and radial geodesic is
\begin{equation}\label{e:rdot}
 \dot{r}=
 \left\{
 \begin{array}{cl}
  -\frac{1-r^2\sin^2\th\Om^2}{1+r^2\sin^2\th\Om^2},
 \\
 1
 \end{array}
 \right.
\end{equation}
The upper (lower) solution represents the ingoing (outgoing) mode.
This is consistent with the fact that the current is null: $J^2=0$,
for the monopole solution, which means that the charges travel
outwards with the speed of light, probably as a result from the
force-free condition.

Consider the charged particle pairs are created outside the LC at
$r=r_{out}$. The s-wave positive energy $\om$ particles cross the LC
inward to $r=r_{in}$ inside the LC. Then the imaginary part of the
action is
\begin{equation}
 \textrm{Im }S=\textrm{Im}\int_{r_{out}}^{r_{in}} p_rdr=-\textrm{Im }
\om\int_{r_{out}}^{r_{in}}
\frac{1+r^2\sin^2\th\Om^2}{1-r^2\sin^2\th\Om^2}dr=\frac{\pi
\om}{\Om\sin\th}.
\end{equation}
Thus, the radiation is thermal with the temperature
\begin{equation}
 T=\hbar\nu \sin\th,
\end{equation}
where $\nu=\Om/2\pi$ is the rotation frequency. So the temperature
is proportional to the rotation frequency of the magnetosphere on
the compact object, which implies that it should be the rotation of
the magnetosphere that provides the energy of radiation. For
millisecond pulsars, the temperature is $\sim
7.6\times10^{-9}\times\nu/(10^3 s^{-1}) K$.

This is the radiation observed for an observer at the center $r=0$.
On the other side (outside the LC), radiation of charged particles
should also be observed. The radiation comes from the particle pairs
created from vacuum. When there is a net flux of radiation towards
$r=0$, there should be a net flux towards $r\rightarrow\infty$.
Unlike the case for a spacetime horizon, this negative-energy
particle flux beyond the LC can be ``seen" since the LC is just a
horizon for the charged particles, but not for the observers. In
black holes, the process that the negative-energy particles fall
into the horizon and reduce their masses remains a mysterious part
of the Hawking radiation theory. The analogous process here is also
puzzling since we do not expect to observe particles with negative
energies at large $r$.

In \cite{2021PhRvD.104b3009L}, we observe that negative energy
indeed can exist in a rotational magnetosphere system in the
Minkowski spacetime. The negative energy being dissipated means that
the thermal energy is extracted. The emergency of the negative
energy is probably due to the observational effect in different
frames. It disappears in the co-rotating frame. We guess that here
the negative energy of the particles outside the LC could also be
taken as an observational phenomenon. These particles could reduce
the rotational energy of the magnetosphere via interaction with the
latter. They might become normal particles with positive energies at
large $r$ by attaining energy from the magnetosphere. But the exact
process needs further detailed study.

As the radiation proceeds, the rotation speed of the magnetosphere
slows down and so the LC expands. The temperature decreases
meanwhile.

\section{Discussion}
\label{sec:discussion}

We have shown that Hawking-like radiation of charged particles could
arise from the LC of a rotating magnetosphere. The temperature is
proportional to the rotation frequency. It is quite low even for the
most rapidly rotating neutron stars. But, this may be theoretically
interesting because this provides the possibility to study the
analogous Hawking radiation process in an electrodynamic system
without any ambiguity.

A possible problem for the temperature is that it is not constant at
different angle $\th$ along the LC. As learnt in gravitational
theory, a horizon must have constant surface gravity or temperature,
satisfying the zero-th law. We think that the difference here may
arise from the fact that the charged particles are restricted to
propagate only along the magnetic field lines. So the temperature
discrepancy on different lines can not be easily erased since the
trans-line communication of charged particles is inhibited. This is
like the situation of sunspots, which can have lower temperatures
than their surrounding photosphere in the presence of magnetic
structures.

The temperature is irrelevant to the strength of the electromagnetic
fields in the magnetosphere. But the field strength should still
play important role. First, a stronger magnetic field is necessary
for restricting higher energy charged particles to move along the
field lines. Second, the field strength exceeding the Schwinger
limit may enhance the radiation because more charged particles can
be excited and created within a duration. Then more particles
possibly cross or tunnel through the LC. But here we do not know how
to reconcile the Schwinger mechanism with the tunneling scenario.

Our discussion above is for the force-free electromagnetic fields in
a charge-filled magnetosphere. For vacuum fields, the charge and
current $I$ are both zero. For example, for the vacuum solution
$\psi=Bz$ with constant and parallel magnetic fields extending in
the $x$ direction in the cylinder coordinates $(x=r\sin\th,
z=r\cos\th)$, the metric is $ds^2=-(1-x^2\Om^2)dt^2+dx^2$. In this
case, the above calculation method is invalid. But this does not
mean that there is no radiation of charged particles since the LC
still exists and the tunneling can still happen. The radiation in
this case is suitable to be tested in laboratory. We may test the
radiation by simply rotating a magnet rapidly.

Finally, it is worth mentioning that the radiation should be not due
to the Unruh effect. It applies only for charged particles. There is
no clue indicating that the rotational acceleration can lead to
thermal creation of particles
\cite{Davies:1996ks,Sriramkumar:1999nw}. The radiation should be due
to the existence of the LC or horizon for the charged particles.


\bibliographystyle{JHEP}
\bibliography{b}

\end{document}